\documentclass[prd,aps,twocolumn,reprint,floatfix,nofootinbib,superscriptaddress]{revtex4}
\pdfoutput=1
\usepackage{amssymb,graphicx}
\usepackage{amsmath}
\usepackage{multirow}
\usepackage{epsfig}
\usepackage[usenames]{color} 
\usepackage[export]{adjustbox}
\usepackage{mathtools}

\newcommand{\beqn}{\begin{eqnarray}}
\newcommand{\eeqn}{\end{eqnarray}}
\newcommand{\beq}{\begin{equation}}
\newcommand{\eeq}{\end{equation}}

\def\tg{\tilde{g}}
\def\tX{\tilde{X}}
\def\tK{\tilde{K}}
\def\tR{\tilde{R}}
\def\tF{\tilde{F}}
\def\tx{\tilde{x}}
\def\tchi{\tilde{\chi}}
\def\tphi{\tilde{\phi}}

\def\AX{\tilde{A}_X}
\def\AXb{\bar{A}_X}
\def\Ap{\tilde{A}_{\partial}}

\def\Apsi{\tilde{A}_\psi}
\def\Lpsi{\tilde{\mathcal{L}}_\psi}
\def\tBox{\tilde{\Box}}
\def\tnabla{\tilde{\nabla}}

\begin{document}

\title{Jordan frame beyond scalar-tensor theories}
\author{Fethi M.\ Ramazano\u{g}lu}
\affiliation{Department of Physics, Ko\c{c} University, \\
Rumelifeneri Yolu, 34450 Sariyer, Istanbul, Turkey }
\date{\today}

\begin{abstract}
We study the Jordan frame formulation of generalizations of scalar-tensor
theories conceived by replacing the scalar with other fields such as vectors.
The generic theory in this family contains higher order
time derivative terms in the Jordan frame action which is indicative of
ill-posedness. However, we show that equations of motion can always be
reduced to a second-order-in-time form as long as the original Einstein frame
formulation is well posed. The inverse transformation
from the Jordan frame back to the Einstein frame is not possible 
for all field values in all theories, but we obtain a fully invertible transformation 
for vector-tensor theories by a redefinition of the vector field. Our main motivation is a better
understanding of spontaneous scalarization and its generalizations,
however our conclusions are applicable to a wide class of theories. Jordan frame
has been traditionally used for certain calculations in scalar-tensor theories of
gravitation, and our results will help researchers generalize these results,
enabling comparison to observational data.
\end{abstract}
\maketitle

\section{Introduction}
Scalar-tensor theories (STTs) have been among the most popular alternatives to general
relativity (GR), and also had a large impact on cosmology~\cite{Fujii:2003pa}. These
theories commonly posit that gravitation is governed by scalar
degrees of freedom in addition to the usual metric of general relativity, 
but their phenomenology can be very diverse otherwise due to various different
coupling terms in their actions. An important feature of STTs
is the freedom to choose the fundamental field variables while formulating them, e.g.
one is always free to redefine a new metric $\tg_{\mu\nu}$ by scaling a given metric
$g_{\mu\nu}$ by a function of the scalar $\phi$: $\tg_{\mu\nu}= A^2(\phi) g_{\mu\nu}$.
Possibilities in such redefinitions are infinite, but two specific cases, called frames,
have been of special importance. The first is the Jordan frame where
the fundamental metric field of the theory couples minimally to matter degrees of freedom,
and the second is the Einstein frame where the metric is such that the metric action
is in the Einstein-Hilbert form, hence identical to that of GR~\cite{Flanagan:2004bz}.

Einstein and Jordan frames have been investigated in great detail in the literature
which has shown their equivalence in many cases~\cite{Flanagan:2004bz},
revealed that one frame can be more useful for analyzing specific
problems such as approximation schemes~\cite{Fujii:2003pa,Damour:1992we,PhysRevD.94.104063,2013PhRvD..87h1506B},
and even led to the discovery of previously overlooked STTs~\cite{Zumalacarregui:2013pma}.
The aim of this work is generalizing the analysis of the relationship between these
two frames to theories that contain higher spin fields such as vectors instead of 
scalars, or less common conformal scaling functions $A(\phi)$ such as those that depend on field derivatives.

Our main motivation is the recently investigated phenomenon of spontaneous tensorization
which is a generalization of spontaneous scalarization in the scalar tensor theories introduced by 
Damour and Esposito-Far{\`e}se (DEF)~\cite{PhysRevLett.70.2220}.
In DEF theories, the scalar fields spontaneously grow to large values from arbitrarily small
perturbations near neutron stars due to a tachyonic instability. Such a theory, with some minor
caveats, confirms to all known weak-field tests while providing large deviations from GR in
the strong field, hence provides an especially good target to be tested using gravitational
wave observations~\cite{PhysRevLett.116.061102,TheLIGOScientific:2017qsa}.
The desirable controlled
spontaneous growth in DEF theories is not a direct results of the scalar nature of the
coupling, or the tachyonic nature of the instability. Any field that carries an
instability, such as a ghost on a vector field, in principle can lead to similar spontaneous
growth which is called spontaneous tensorization~\cite{Ramazanoglu:2017xbl,Ramazanoglu:2017yun}.
Overall, the theory of DEF is but one member of a large family of theories with similar observational
signatures, all of which can be potentially tested with gravitational waves in the near
future~\cite{Ramazanoglu:2017yun}.

All spontaneous tensorization theories have been formulated in the Einstein frame for reasons 
we will discuss, and he main theme of this study is their properties in the Jordan frame.
Despite our motivation, we
will not specify our coupling terms to those that incite spontaneous growth, hence our
results are general. We will use the terminology of Einstein and Jordan
frames in a generalized sense, the former is always the one where the gravitational action is in the
Einstein-Hilbert form, and the latter is always the one where matter fields couple to the metric minimally.

In Sec.~\ref{sec:scalar_tensor} we present the tranformation between
Einstein and Jordan frames in the quintessential STT
of Brans and Dicke (BD)~\cite{PhysRev.124.925} whose conformal matter coupling structure is kept
in all other theories we are interested. In Sec.~\ref{sec:VectorTachyon}
we obtain the Jordan frame for a vector-tensor theory. In Sec.~\ref{sec:scalar_ghost}
we go back to scalar fields, but this time study derivative couplings.
We demonstrate the existence of higher derivative terms in the Jordan frame,
commonly indicative of ill-posedness, and present the results from the existing
literature which resolve this problem. We also discuss the invertibility of the frame transformations.
In Sec.~\ref{sec:vector_ghost} we analyze the most general
spontaneous tensorization theory which also has potentially dangerous higher
derivative terms in the Jordan frame. We address this problem by showing that the equations of motion have at
most second order time derivatives. In the last section, we summarize and discuss our results.

\section{Changing the frame in Scalar-Tensor Theories}\label{sec:scalar_tensor}
The most elementary case to compare the Einstein and Jordan frames is
the Brans Dicke theory which was first introduced in the Jordan frame~\cite{PhysRev.124.925}
\begin{align}\label{st_action_jordan}
 \frac{1}{16\pi} &\int d^4x \sqrt{-\tilde{g}}\ \Phi \tilde{R} -  \frac{1}{16\pi} \int d^4x \sqrt{-\tilde{g}}\
\frac{\omega(\Phi)}{\Phi} \tilde{g}^{\mu \nu} \partial_{\mu} \Phi  \partial_{\nu} \Phi
  \nonumber \\
 &+ S_m \left[f_m, \tilde{g}_{\mu \nu} \right]
\end{align}
where $\Phi$ is the scalar field, $\tilde{R}$ is the Ricci scalar associated with the
Jordan frame metric $\tilde{g}_{\mu\nu}$, 
and $f_m$ denotes any matter degrees of freedom in the spacetime with their respective
action $S_M$. $\omega = constant$ for BD, but it can be generalized to obtain other 
STTs, for example $\omega = -3/2-1/(2\beta \log \Phi)$ with a
negative constant $\beta$ gives the spontaneous scalarization of DEF~\cite{2013PhRvD..87h1506B}. 

Eq.~\ref{st_action_jordan} deviates from the action of
a scalar field minimally coupled to GR by the $\Phi$ factor in the Einstein-Hilbert-like first term, and by the
non-canonical scalar field action. The former issue can be addressed by expressing the action in
terms of another metric which is conformally related the the original, $g_{\mu\nu}= A^{-2}\tg_{\mu \nu}$,
since then
\begin{align}\label{ricci_identity}
 \tilde{R} &= A^{-2} R - 6 g^{\mu\nu} A^{-3} \nabla_\mu \nabla_\nu A \nonumber \\
 \Leftrightarrow R &= A^{2} \tR + 6 \tg^{\mu\nu} A^{-1} \tnabla_\mu \tnabla_\nu A 
\end{align}
which can be shown by straightforward calculation of the Ricci scalar
in 4 spacetime dimensions. Here, all quantities with a tilde are related to
$\tg_{\mu\nu}$, and bare ones are related to $g_{\mu\nu}$.\footnote{
Our tilde convention follows DEF and the spontaneous scalarization/tensorization
literature. The opposite is sometimes employed e.g. in~\cite{Zumalacarregui:2013pma}.}
Remembering that $\tg = A^4 g$ and $\tg^{\mu\nu}=A^{-2} g^{\mu\nu}$, we immediately see that the choice 
$\tilde{g}_{\mu \nu}= \Phi^{-1}g_{\mu \nu}$, puts the action in the form
\begin{align}\label{st_action_inter}
 \frac{1}{16\pi} &\int d^4x \sqrt{-g}\ \left[ R 
 -\frac{3/2}{\Phi^2} g^{\mu \nu} \partial_{\mu} \Phi  \partial_{\nu} \Phi \right]
 \nonumber \\
 -  \frac{1}{16\pi} & \int d^4x \sqrt{-g}\
\frac{\omega(\Phi)}{\Phi^2} g^{\mu \nu} \partial_{\mu} \Phi  \partial_{\nu} \Phi
  \nonumber \\
 &+ S_m \left[f_m, \Phi^{-1}g_{\mu \nu} \right]
\end{align}
up to boundary terms. The first term is exactly that of GR as we desired.

The second problem of having a non-canonical scalar action can also be addressed
by using our freedom to redefine the scalar field. Introducing $\phi$ such that
\begin{align}\label{BD_redefinition}
\frac{d\Phi}{d\phi} &=\sqrt{\frac{4\Phi^2}{3+2\ \omega(\Phi)}}
\end{align}
finally puts the action into the commonly used Einstein frame form
\begin{align}\label{st_action_einstein}
 \frac{1}{16\pi} &\int d^4x \sqrt{-g}\   R -   \frac{1}{16\pi} \int d^4x \sqrt{-g}\  
2g^{\mu \nu} \partial_{\mu} \phi  \partial_{\nu} \phi
  \nonumber \\
 &+ S_m \left[f_m, A^2(\phi) g_{\mu \nu} \right] \ .
\end{align}
We call $\ A(\phi) \equiv \left(\Phi(\phi)\right)^{-1/2}$ the conformal scaling function.
 All the ``alternative''
nature of this action is in its matter coupling, the first two
terms simply represent a scalar field living under GR. The price to have these familiar
action terms is the nonminimal matter coupling.
For the BD and
DEF theories, the scalar field redefinitions are
\begin{align}\label{scalar_A_def}
\Phi = \exp\left(\sqrt{\frac{4}{3+2\omega}} \phi \right) \ , \ 
\Phi = \exp\left(-\beta \phi^2 \right)
\end{align}
respectively. 

\section{Jordan Frame in Theories of Spontaneous Vectorization}\label{sec:VectorTachyon}
A close examination of the DEF theory in Eq.~\ref{st_action_einstein}
shows that one can get similar alternative theories of gravity by replacing the 
scalar by a vector as~\cite{BeltranJimenez:2013fca,Ramazanoglu:2017xbl}
\begin{align}\label{vt_action_einstein}
 \frac{1}{16\pi} &\int d^4x \sqrt{-g}\ R 
 -  \frac{1}{16\pi} \int d^4x \sqrt{-g}\ g^{\mu\rho} g^{\nu\sigma} F_{\rho\sigma} F_{\mu\nu} \nonumber \\
 &+ S_m \left[f_m, A_X^2(x) g_{\mu \nu} \right], \ x =g^{\mu\nu}X_\mu X_\nu \ ,
\end{align}
where $X_\mu$ is a vector field that governs gravity in addition to the metric, similarly to the scalar field
in STTs, and 
$F_{\mu\nu} = \nabla_\mu X_\nu -\nabla_\nu X_\mu= \partial_\mu X_\nu -\partial_\nu X_\mu$.
We keep all vector-related quantities in the lower index to explicitly see the
the inverse metric terms.

It is clear from this presentation why Einstein frame is more suitable to generalize
STTs to vectors or other fields. Both the metric (Einstein-Hilbert) and the scalar
field actions are in their ``standard'' forms, hence one keeps the Einstein-Hilbert action and replaces the
action of the scalar with the standard action for a vector field to obtain a ``vector-tensor''
theory.\footnote{In the original formulation
the vector is massive to conform to some possible observational bounds~\cite{Ramazanoglu:2017xbl}.
However, intrinsic mass of 
vectors or scalars do not affect the current discussion.} Whereas, it is hard to see how to
change the unusual scalar field terms in the Jordan frame of Eq.~\ref{st_action_jordan}
to those of vectors. 
In Eq.~\ref {vt_action_einstein}, modification to GR comes in the form of a confomal
scaling of the metric that interacts with matter fields, just as for the DEF or BD theories.
Then our only choice is the function $A_X$. It has been shown that 
$A_X=e^{\beta_X x/2}$  with a constant $\beta_X$
or a similar function whose leading behavior around 0 is second order in
its argument $x$ leads to spontaneously growing vector fields in the vicinity of 
neutron stars in analogy to spontaneous scalarization.

Remembering that we define the Jordan frame of a theory as
where the metric couples minimally to the matter fields, we want 
to express the action in terms of $\tilde{g}_{\mu\nu} = A_X^2 \ g_{\mu\nu} $
for which the action becomes
\begin{align}\label{vt_action_jordan}
 \frac{1}{16\pi} &\int d^4x \sqrt{-\tilde{g}}\ \AXb^{-2} \tR 
 -\frac{1}{16\pi} \int d^4x \sqrt{-\tg}\ \tg^{\mu\rho} \tg^{\nu\sigma} F_{\rho\sigma} F_{\mu\nu}  \nonumber \\
 +\frac{6}{16\pi}& \int d^4x \sqrt{-\tg}\ \tg^{\mu\nu} \partial_\mu \AXb^{-1} \partial_\nu \AXb^{-1}  \nonumber \\
 &+ S_m \left[f_m, \tg_{\mu \nu} \right]\ .
\end{align}
$\AXb$ is simply $A_X$, but now implicitly defined as a function of the
field variables in the new frame by
\begin{align}\label{ax_implicit}
\AXb(\tg^{\mu\nu}X_\mu X_\nu) = A_X(\AXb^2  \tg^{\mu\nu}X_\mu X_\nu )\ .
\end{align} 
For example, the exponential function we used before now means
\begin{align}\label{ax_implicit2}
\AXb^2 = e^{\beta_X \tg^{\mu\nu}X_\mu X_\nu  \AXb^2 }\ .
\end{align} 

The implicit definition in Eq.~\ref{ax_implicit} is unappealing, and more importantly
cannot be always inverted for given $\{\tg_{\mu\nu}, X_\mu\}$, rendering the theory
meaningless for such values in the Jordan frame. We will discuss
this issue in more detail in the coming sections, but we can
address it for this specific theory by utilizing the freedom to redefine
the vector field. Consider
\begin{align}\label{X_redefine}
\tX_\mu \equiv \bar{A}_X X_\mu \ .
\end{align}
Now we can re-express $A_X$ as a function of $\tg_{\mu\nu}$ and $\tX_\mu$ as
\begin{align}\label{ax_explicit}
\AX(\tg^{\mu\nu} \tX_\mu \tX_\nu) = A_X(\tg^{\mu\nu} \tX_\mu \tX_\nu )\ ,
\end{align} 
that is $\AX$ has the exact same functional form as $A_X$, but for a different
set of variables. For the exponential function we considered before
\begin{align}\label{ax_explicit}
\AX = e^{\beta_X \tg^{\mu\nu}\tX_\mu \tX_\nu/2}\ .
\end{align} 
Thus, one can transform from one frame and set of field variables to the other in the
straightforward manner
\begin{align}\label{vt_transform}
\tg_{\mu\nu}& =A_X^{2}  g_{\mu\nu} \ \  &g_{\mu\nu}& = \AX^{-2}\ \tg_{\mu\nu} \nonumber \\
\tX_\mu& =A_X  X_\mu  \ \ &X_\mu& = \AX^{-1} \tX_\mu \\
A_X&= e^{\beta_X g^{\mu\nu}X_\mu X_\nu/2} \ \  
&\AX &= e^{\beta_X \tg^{\mu\nu}\tX_\mu \tX_\nu/2} \nonumber 
\end{align}
The transformation $\{g_{\mu\nu}, X_\mu\}\leftrightarrow \{\tg_{\mu\nu}, \tX_\mu\}$
is invertible. We will see that this is not always the case in other generalizations 
of STTs.

Finally, the Jordan frame action in terms of the new variables
$\{\tg_{\mu\nu}, \tX_\mu \}$ is given by
\begin{align}\label{vt_action_jordan3}
 \frac{1}{16\pi} &\int d^4x \sqrt{-\tilde{g}}\ \AX^{-2} \tR 
 -\frac{1}{16\pi} \int d^4x \sqrt{-\tg}\ \AX^{-2} \tF^{\mu\nu}\tF_{\mu\nu}  \nonumber \\
 -\frac{2}{16\pi}& \int d^4x \sqrt{-\tg}\ \AX^{-2} \tilde{\Lambda}_X^2
\bigg[ (\tx-3-\tilde{\Lambda}_X^{-1}) \tnabla^\mu \tx \tnabla_\mu \tx 
   \nonumber \\
&-  (\tX^\mu \tnabla_\mu \tx)^2  +2\tilde{\Lambda}_X^{-1}   (\tX^\rho \tnabla_\rho\tX^\mu)\tnabla_\mu \tx
\bigg]\nonumber \\
 &+ S_m \left[f_m, \tg_{\mu \nu} \right] \ ,
\end{align}
where $\tilde{F}_{\mu\nu} \equiv \tilde{\nabla}_\mu \tX_\nu -\tilde{\nabla}_\nu \tX_\mu$,
$\tx \equiv \tg^{\mu\nu} \tX_\mu \tX_\nu$, $\tilde{\Lambda}_X \equiv  \AX^{-1} (d\AX/d\tx)$ and
all raising is performed with $\tg^{\mu\nu}$.
The action for the vector field is not the
standard one, but this is expected in the Jordan frame as in Eq.~\ref{st_action_jordan}.

We should note that the actions in 
Eq.~\ref{vt_action_jordan} and Eq.~\ref{vt_action_jordan3} both satisfy 
our definition of the Jordan frame since they both have minimal matter
coupling. However, we will prefer the latter since it does not suffer from 
the nonexistence of $\bar{A}_X$ for certain field values in 
Eqs.~\ref{ax_implicit},~\ref{ax_implicit2}, hence we 
consider $\{\tg_{\mu\nu}, \tX_\mu\}$ to provide a more natural setting for
the Jordan frame.

$\tX$ carries three degrees of freedom rather than the usual two
in Maxwell fields since it lacks the gauge freedom $\tX_\mu \to \tX_\mu + \partial_\mu \lambda$.
We can consider the degree of freedom in the norm $\tx$
separately, and view $\tX_\mu$ as carrying the remaining two.
Thus, we can treat $\tx$ as an independent scalar which couples to
the vector action $\tF^{\mu\nu}\tF_{\mu\nu}$ in addition to the gravity term $\tR$.
This is reminiscent of Einstein-Maxwell-scalar theories, especially the ones
that feature a newly discovered type of spontaneous scalarization~\cite{Herdeiro:2018wub}. 
This might lead to an alternative understanding of the vectorization process through the
scalar degree of freedom in $\tx$, and provide new connections between spontaneous
vectorization and spontaneous scalarization. Note that one can also define
$\tchi =\AX^{-2}$ and express Eq.~\ref{vt_action_jordan3} in
somewhat closer resemblance to Eq.~\ref{st_action_jordan} where 
$\tchi$ would behave like a scalar. 

\section{Jordan frame for ghost-based spontaneous scalarization}\label{sec:scalar_ghost}
A second avenue to generalize spontaneous growth of DEF is
using a different instability, as opposed to using a different field.
More concretely, the tachyonic can be replaced
by a ghost if the conformal scaling depends on the derivative
terms in the action
\begin{align}\label{st_action_ghost_einstein}
 \frac{1}{16\pi} &\int d^4x \sqrt{-g}\   R -   \frac{1}{16\pi} \int d^4x \sqrt{-g}\  
2g^{\mu \nu} \partial_{\mu} \phi  \partial_{\nu} \phi \\
 &+ S_m \left[f_m, A_{\partial}^2(K) g_{\mu \nu} \right] \ , \ 
 K \equiv g^{\mu\nu}\partial_\mu \phi \partial_\nu \phi \ . \nonumber 
\end{align}
The resulting theory leads to scalarization of neutron stars, e.g
for the choice $A_\partial(K)=e^{\beta_\partial K/2}$ with some
constant $\beta_\partial$~\cite{Ramazanoglu:2017yun}. This theory is named
ghost-based spontaneous scalarization since it can be shown that small 
perturbations around the scalar field vacuum behave like a
ghost, but this instability is suppressed as the field grows.

Using Eq.~\ref{ricci_identity}, we can transform to the Jordan frame with
$\tg_{\mu\nu} = A_\partial^2 g_{\mu\nu}$
\begin{align}\label{st_action_ghost_jordan}
 \frac{1}{16\pi} &\int d^4x \sqrt{-\tilde{g}}\ \Ap^{-2} \tR  -   \frac{2}{16\pi} \int d^4x \sqrt{-\tg}\  \Ap^{-2} \tK \nonumber  \\
+\frac{6}{16\pi}& \int d^4x \sqrt{-\tg}\ \tg^{\mu\nu} \partial_\mu \Ap^{-1} \partial_\nu \Ap^{-1}  \nonumber \\
 &+ S_m \left[f_m, \tg_{\mu \nu} \right]  \ , \  
 \tK \equiv \tg^{\mu\nu}\partial_\mu \phi \partial_\nu \phi  \ . 
\end{align}
$\Ap(\tK)$ is defined implicitly through
\begin{align}\label{ap_implicit}
\Ap(\tK)= A_\partial(\Ap^2(\tK)\tK)\ ,
\end{align}
e.g. the exponential conformal scaling above leads
to\footnote{ We can have an explicit definition in this specific case in terms
of the relatively well-known Lambert $W$ function $W(xe^x) \equiv x$: 
$\Ap(\tK)=\sqrt{W(-\beta_\partial \tK)/(-\beta_\partial \tK )}$. However,
the definition is implicit for generic $A_\partial$.}
\begin{align}\label{ap_implicit2}
\Ap^2= e^{\beta_\partial \tK \Ap^2}\ . 
\end{align}

Even though this Jordan frame formulation looks quite similar to that of the vector field case
in Eq.~\ref{vt_action_jordan} and~\ref{ax_implicit}, it brings in new issues to tackle. First,
Eq.~\ref{st_action_ghost_jordan} contains terms with more than one time derivatives of $\phi$
through $\partial_\mu \Ap$, which might indicate an unphysical nature due to Ostrogradsky's
theorem~\cite{Woodard:2015}. 
This is peculiar, since we do not expect the nature of the theory to change radically 
from one frame to the other, and there are no higher time derivative terms in the
Einstein frame formulation in Eq.~\ref{st_action_ghost_einstein}.

This puzzle can be resolved by considering the equations of motion in the
Jordan frame~\cite{Zumalacarregui:2013pma}
\begin{align}\label{eom_ghost_jordan}
 \tilde{G}_{\mu\nu} &= 8\pi\Ap^{2}   \tilde{T}_{\mu\nu} 
-2\Ap (\tg_{\mu\nu} \tBox\Ap^{-1}-\tnabla_\mu \tnabla_\nu \Ap^{-1} ) \\
&-2\tilde{\Lambda}_\partial (6\Ap\tBox \Ap^{-1} - \tR)  \partial_\mu \phi \partial_\nu \phi \nonumber \\
&+\Ap^{-2} \tg_{\mu\nu}  \tg^{\rho\sigma}\partial_\sigma \Ap \partial_\rho \Ap 
-4\Ap^{-2} \partial_\mu \Ap \partial_\nu \Ap \nonumber \\
&+ 2\partial_\mu \phi \partial_\nu \phi - \tg_{\mu\nu} \tK
-4\tilde{\Lambda}_\partial \tK \partial_\mu \phi \partial_\nu \phi \nonumber\\
0&=\tnabla_\mu \left( \tilde{\Lambda}_\partial \Ap^{-2}
(6\Ap\tBox \Ap^{-1} - \tR+2\tK-\tilde{\Lambda}_\partial^{-1}) \tnabla^\mu \phi \right)
\nonumber
\end{align} 
where $\tilde{\Lambda}_\partial \equiv \Ap^{-1} (d\Ap/d\tK)$.
These equations indeed have up to fourth order time derivatives indicative of ill-posedness.
However, the trace of the first equation gives
\begin{align}\label{eom_ghost_jordan_trace}
6\Ap\tBox \Ap^{-1} - \tR = 
\frac{8\pi\Ap^{2} \tilde{T}-2\tK-4\tilde{\Lambda}_\partial \tK^2}{1+2\tilde{\Lambda}_\partial \tK} \ .
\end{align}  
Hence, as long as the stress energy tensor of matter
is first order, $6\Ap\tBox \Ap^{-1} - \tR$ can be re-expressed in terms of 
first derivatives, and $\partial_t^2 \Ap$ can be re-expressed
in terms of at most second time derivatives of $\phi$ and $\tg_{\mu\nu}$.
Inserting these identities back ensures that Eq.~\ref{eom_ghost_jordan}
contains at most second time derivatives~\cite{Zumalacarregui:2013pma}.

The second issue in the Jordan frame formulation is the implicit definition of
$\Ap$ in Eq.~\ref{ap_implicit} which does not necessarily have a solution for all
$\tK$ for a given $A_\partial$. For example, the exponential function
we discussed (Eq.~\ref{ap_implicit2}) only provides a solution for $\Ap$ if
$\beta_\partial\tK<e^{-1}$ and is multi-valued for positive $\beta_\partial \tK$. This
also implies that it is not always possible to invert $\tg_{\mu\nu}$ to
$g_{\mu\nu}$. In some sense, all possible values of
$\{\tg_{\mu\nu},\phi \}$ is too big a configuration space for our theory.
This is not the case for all $\Ap$, for example the choice
$A_\partial(K)=e^{\gamma_\partial K^2}$ leads to a $\Ap(\tK)$ defined
for all $\tK$ values~\cite{Zumalacarregui:2013pma}. 

Non-invertibility of the frame transformation is not a new problem, and
is present in the simplest STTs such as BD as well. In Eq.~\ref{st_action_jordan},
$\Phi$ is constrained to be positive. Field variables cannot be transformed
to the Jordan frame otherwise due to Eq.~\ref{scalar_A_def}, and the action also
becomes meaningless for the
DEF theory due to the logarithmic terms in $\omega(\Phi)$.
Nevertheless, we would still want to have an invertible transformation
between frames for as much of the configuration space as possible.

Note that we encountered the non-invertibility problem in vector-tensor theories as well
(Eq.~\ref{ax_implicit},~\ref{ax_implicit2}), but the redefinition of the vector field,
$X_\mu \to \tX_\mu$ resolved the issue by providing a fully invertible
transformation between the frames in Eq.~\ref{vt_transform}.
In other words, the full range of $\tX_\mu$ was just right for our theory while
that of $X_\mu$ was ``too big''.
We were not able to construct any such redefinition $\phi \to \tphi$
in the current case. As an example, the most obvious candidate
\begin{align}
\partial_\mu \tphi \equiv \Ap \partial_\mu \phi
\end{align}
leads to the explicit definition 
$\Ap= A_\partial(\tg^{\mu\nu}\partial_\mu \tphi \partial_\nu \tphi )$ and a fully invertible
tranformation between frames. However,
this also implies
$\partial_\mu \partial_\nu \tphi \neq \partial_\nu \partial_\mu \tphi$, hence
such a differentiable $\tphi$ does not exist. 

A related problem to the nonexistence of $\Ap$ for certain field values would be the
following: if we start with some initial data $\{\tg_{\mu\nu},\phi \}$ and their
first derivatives in the Jordan frame for which $\Ap$ is defined, does the time
evolution lead to values of $\{\tg_{\mu\nu},\phi \}$ for which $\Ap$ is defined
everywhere in the future of the initial data? Since the non-existence of $A_\partial$ is not a problem in
the Einstein frame for any values of $\{g_{\mu\nu},\phi \}$, we would expect
the answer to be affirmative. A proper analysis requires tools from the theory
of partial differential equations, and we hope mathematical physics can provide
some insight for this problem which has not been addressed in the gravitational
physics literature to the best of our knowledge.

Lastly, the fact that the equations of motion are ultimately second order in time
derivatives in the Jordan frame may suggests that there
is also a field redefinition that would allow the action to contain only first order 
derivatives, but we could not identify a simple example of this aside from the
trivial transformation of going back to the Einstein frame.

\section{Jordan frame for generic spontaneous tensorization}\label{sec:vector_ghost}
We have seen that spontaneous growth can be generalized from scalars to vectors,
or from a tachyon-based mechanism to a ghost-based mechanism. 
This approach can be continued to various other
fields  such as spinors~\cite{Ramazanoglu:2018hwk}, other mechanisms such as
spontaneous growth through the Higgs mechanism~\cite{Ramazanoglu:2018tig},
or any combination of them. All these form the family of spontaneous tensorization
theories. 

How generic are the issues of higher time derivatives and
invertibility of frame transformations that we encountered in Sec.~\ref{sec:scalar_ghost}?
If they appear in a generic spontaneous tensorization theory, can
they be resolved similarly to the case of ghost-based spontaneous 
scalarization? Consider the following general action for which all the theories we
have investigated so far are special cases
\begin{align}\label{psi_action_einstein}
 \frac{1}{16\pi} &\int d^4x \sqrt{-g}\   R -   \frac{1}{16\pi} \int d^4x \sqrt{-g}\  
\mathcal{L}_\psi(g^{\mu\nu},\psi, \partial\psi)
  \nonumber \\
 &+ S_m \left[f_m, A_\psi^2(g^{\mu\nu},\psi, \partial\psi) g_{\mu \nu} \right] \ .
\end{align}
Here, $\psi$ is any field, potentially with multiple tensor indices, and $\partial \psi$
is a collective symbol for its first derivatives. The Jordan
frame of such a theory is obtained by $\tg_{\mu\nu}= A_\psi^2 g_{\mu\nu}$ which has the action
\begin{align}\label{psi_action_jordan}
 \frac{1}{16\pi} &\int d^4x \sqrt{-\tilde{g}}\ \Apsi^{-2} \tR 
 -   \frac{1}{16\pi} \int d^4x \sqrt{-\tg}\ \Lpsi(\tg^{\mu\nu},\psi, \partial\psi)  \nonumber \\
 +\frac{6}{16\pi}& \int d^4x \sqrt{-\tg}\  \tg^{\mu\nu}
  \partial_\mu \Apsi^{-1} \partial_\nu \Apsi^{-1} \nonumber \\
 &+ S_m \left[f_m, \tg_{\mu \nu} \right]  \ . 
\end{align}
where $\Apsi(\tg^{\mu\nu},\psi, \partial\psi)$ is defined implicitly as 
\begin{align}\label{apsi_implicit}
\Apsi = A_\psi(\Apsi^2\ \tg^{\mu\nu},\psi, \partial\psi) \ ,
\end{align} 
and 
\begin{align}\label{apsi_implicit}
\Lpsi(\tg^{\mu\nu},\psi, \partial\psi) = \Apsi^{-4}\ \mathcal{L}_\psi(\Apsi^2\ \tg^{\mu\nu},\psi, \partial\psi) \ .
\end{align} 
The crucial point is that, $\Lpsi$ and $\Apsi$
are still only functions of at most the first derivatives of $\psi$.

The equations of motion in the Jordan frame are
\begin{align}\label{eom_psi_jordan}
 \tilde{G}_{\mu\nu} &= 8\pi\Apsi^{2}   \tilde{T}_{\mu\nu} 
-2\Apsi (\tg_{\mu\nu} \tBox\Apsi^{-1}-\tnabla_\mu \tnabla_\nu \Apsi^{-1} ) \\
&-2\ (6\Apsi \tBox \Apsi^{-1} - \tR)\Apsi^{-1} \frac{\delta \Apsi}{\delta \tg^{\mu\nu}} \nonumber \\
&+\Apsi^{-2} \tg_{\mu\nu}  \tg^{\rho\sigma}\partial_\sigma \Apsi \partial_\rho \Apsi 
-4\Apsi^{-2} \partial_\mu \Apsi \partial_\nu \Apsi \nonumber \\
&-\frac{\Apsi^2}{\sqrt{-\tg}}  \frac{\delta (\sqrt{-\tg}\Lpsi)}{\delta \tg^{\mu\nu}}  \nonumber\\
0&=\frac{1}{\sqrt{-\tg}}\partial_\mu \left(  \sqrt{-\tg}\Apsi^{-3}
(6\Apsi\tBox \Apsi^{-1} - \tR) \frac{\delta \Apsi}{\delta (\partial_\mu \psi)} \right) \nonumber \\
&-\Apsi^{-3}(6\Apsi\tBox \Apsi^{-1} - \tR) \frac{\delta \Apsi}{\delta  \psi}-\frac{1}{2} \frac{\delta \Lpsi}{\delta \psi}  \ .
\nonumber
\end{align} 
The second equation contains the fourth derivative term $\tnabla_\mu \tBox \Apsi^{-1}$
and both equations have third time derivatives of $\psi$. Both cases are potentially problematic.
However, the trace of the first equation implies
\begin{align}\label{eom_psi_jordan_trace}
6\Apsi\tBox \Apsi^{-1} - \tR = 
\frac{8\pi \tilde{T}-\frac{\Apsi^2}{\sqrt{-\tg}}  \frac{\delta (\sqrt{-\tg}\Lpsi)}{\delta \tg^{\mu\nu}}
}{1+2\Apsi^{-1} \frac{\delta \Apsi}{\delta \tg^{\mu\nu}} } \ ,
\end{align}  
in complete analogy to Eq.~\ref{eom_ghost_jordan_trace}.
$6\Apsi\tBox \Apsi^{-1} - \tR$ can be expressed in terms of
at most first derivatives of the fundamental field variables $\{\tg_{\mu\nu},\psi\}$
as long as $\Apsi$, $\Lpsi$ and matter stress-energy contain at most first derivatives.
This in turn means $\partial_t^2 \Apsi$ can also be expressed in terms of at most two time derivatives.
Hence, changing from the Einstein frame (Eq.~\ref{psi_action_einstein}),
to the Jordan frame (Eq.~\ref{psi_action_jordan}) does not introduce ill-posedness.

Note that our starting action Eq.~\ref{psi_action_einstein} in the Einstein frame is
general enough to contain potential self interaction terms for $\psi$ in $\mathcal{L}$,
and a conformal scaling $A_\psi$ that depends both on $\psi$ and its derivatives.
Thus, our results continue to hold for the spontaneous growth of massive scalars
or vectors, or theories where ghost-based and tachyon-based instabilities are 
present at the same time. It is also trivial to generalize the result to a collection
if fields $\psi^{(i)}$. These cover all examples of spontaneous tensorization in the
literature~\cite{Ramazanoglu:2017xbl,Ramazanoglu:2017yun,Ramazanoglu:2018hwk,Ramazanoglu:2018tig}. 
All these theories, aside from spontaneous vectorization in Sec.~\ref{sec:VectorTachyon},
also contain conformal scaling functions which contain derivatives.
Hence, the resolution of the higher time derivative problem we outlined
is central to their viability in the Jordan frame as physical theories.

Invertibility of the frame transformation in relation to the existence of a solution to Eq.~\ref{apsi_implicit}
for $\Apsi$ is not resolved in general. All possible values of $\psi$ provide too big
a configuration space in the Jordan frame, and the questions we posed in
Sec.~\ref{sec:scalar_ghost} are open in this generic case as well.
However, we remind that such problems are present even in the DEF theory.

\section{Conclusions}
We studied the Jordan frame formulation of generalizations of
STTs, where the scalar is replaced with other fields, and
couplings can depend on derivatives. Our motivation came from the specific
class of theories that feature spontaneous tensorization. These are most
naturally defined in the Einstein frame where the action for the additional field
to the metric (e.g. the vector) is in the canonical form. However, our results
can be applied to any generalization of STTs that is based on a 
conformal scaling of the metric in the matter action by some function of a dynamical field
and its first derivatives.

The first case we examined is the vector-tensor theory obtained by  replacing the
scalar in the DEF theory by a vector where the conformal scaling function $A_X$
depends on the norm of the vector field. A completely invertible
transformation (Eq.~\ref{vt_transform}) can be obtained if the vector field is redefined in the Jordan frame,
$X_\mu \to \tX_\mu=A_X X_\mu$, as well as the metric. Moreover,
interesting connections can be observed to the recently discovered spontaneous growth
in Einstein-Maxwell-scalar theories~\cite{Herdeiro:2018wub}.

Jordan frame of ghost-based spontaneous scalarization where the conformal 
scaling depends on the derivatives of the scalar field presents challenges.
First, the Jordan frame contains higher derivative terms indicative of ill-posedness due
to Ostrogradsky's theorem, which is odd since this is not an issue in the Einstein frame,
and we would not expect the nature of the theory to change in such radical fashion due
to a frame change. The equations of motion indeed contain up to fourth order time
derivatives, but it can be shown that such terms cancel each other
to render the equations second order in time. It is also curious to see that the conformal
scaling function cannot be defined for all values of the field variables
$\{\tg_{\mu\nu}, \phi\}$ which causes the transformation between frames to be
non-invertible. We could not find a formulation where this problem is resolved, but noted
that this is the case even in the DEF theory where the scalar field is restricted to be
positive in the Jordan frame. It is important to understand 
the meaning of the field values in the Jordan frame where the transformation back to
the Einstein frame is not defined, which we leave to future studies.

We finally showed that a generic spontaneous tensorization theory 
contains higher time derivative  terms in its formulation, much like ghost-based
spontaneous scalarization. The equations of motion are again ultimately rendered 
second order in time, even though they naively contain fourth time derivatives, 
demonstrating that the Jordan frame formulation does not introduce ill-posedness.
We should add at this point that the fact that there are only first derivative terms
in the Einstein frame action does not guarantee well-posedness. A theory can be
rendered unphysical by other factors such as indefinitely growing fields such as
ghosts, even if the equations of motion have no more than two time derivatives.
The Einstein frame formulation of spontaneous tensorization theories are not known to be completely free of 
such undesirable features~\cite{Ramazanoglu:2017yun,Ramazanoglu:2018hwk,Ramazanoglu:2018tig},
but our work here shows that transferring to the Jordan frame at least does not add
new sources of ill-posedness.

Certain calculations on STTs have been performed
using the Jordan frame such as the gravitational wave memory for the DEF
theory~\cite{PhysRevD.94.104063}. Consequently, we believe this study will enable researchers to extend 
similar work to spontaneous tensorization in general. Possibility of near-future
testing is a basic appeal of spontaneous scalarization and tensorization.  Calculations
of specific observational signs will enable the gravity community to compare the predictions
of these theories to actual observations, and understand the differences between
individual theories in the spontaneous tensorization family. 
 
\acknowledgments
This project started with a question posed by Leonardo
Gualtieri to whom we are grateful.
The author is supported by Grant No.~117F295
of the Scientific and Technological Research Council
of Turkey (T{\"U}B{\.I}TAK). We would like to acknowledge
networking and travel support by the COST Action
CA16104.

\bibliographystyle{h-physrev}
\bibliography{spinorization}

\end{document}